\documentclass[]{aa}
\usepackage[utf8]{inputenc}
\usepackage{graphicx}
\usepackage[varg]{txfonts}
\usepackage{hyperref}
\usepackage{amsmath}
\usepackage{natbib}
\usepackage{graphicx}
\usepackage{xcolor}

\newcommand\I{{\mathrm{i}\mkern2mu}}
\newcommand\E{{\mathrm{e}}}
\newcommand\D{\mathrm{d}}
\newcommand\co[1]{{{#1}^*}}
\newcommand\mat{\mathrm}

\begin{document}

\title{The elliptical power law profile lens}

\author{%
Nicolas Tessore\thanks{\email{nicolas.tessore@unibo.it}}
\and
R. Benton Metcalf
}

\institute{%
Department of Physics and Astronomy, Università di Bologna, viale Berti Pichat 6/2, 40127 Bologna, Italy
}

\abstract{%
The deflection, potential, shear and magnification of a gravitational lens following an elliptical power law mass model are investigated.
This mass model is derived from the circular power law profile through a rescaling of the axes, similar to the case of a singular isothermal ellipsoid.
The resulting deflection can be calculated explicitly and given in terms of the Gaussian hypergeometric function.
Analytic expressions for the remaining lensing properties are found as well.
Because the power law profile lens contains a number of well-known lens models as special cases, the equivalence of the new expressions with known results is checked.
Finally, it is shown how these results naturally lead to a fast and accurate numerical scheme for computing the deflection and other lens quantities, making this method a useful tool for realistically modelling observed lenses.
}


\maketitle

\section{Introduction}

Investigations in strong gravitational lensing often require an
approximation of the lens system by an idealised model with a fixed
radial mass profile.
In the usual formulation of gravitational lensing, the three-dimensional mass distribution is integrated along the line of sight, resulting in a two-dimensional surface mass density $\kappa$ that is responsible for the deflection of passing light rays (an introduction and reference can be found e.g. in \citealp*{2006glsw.conf.....M}).
The initial model for a lens system is often a spherically symmetric mass distribution, which leads to a circularly symmetric surface mass density.
Many of the classical lenses fall into this category, such as isothermal spheres or the Navarro, Frenk and White profile \citep*{1996ApJ...462..563N,1996A&A...313..697B}.
Because of the high level of symmetry, their lensing properties can often be worked out analytically.

The next step in approximating real lens systems is to turn the known lenses with spherical symmetry into ellipsoids by rescaling an arbitrary axis of either the surface mass density or the deflection potential, fixing the other through the Poisson equation (e.g. \citealp{1993ApJ...417..450K} and references therein).
Choice of an elliptical potential -- often called a pseudo-elliptical
model -- simplifies the problem, because the
shear, deflection, and convergence can all be expressed directly in terms of derivatives of the potential, thus eliminating the need to solve complicated integrals.
However, this approach can lead to unrealistic and unphysical surface mass densities with peanut-shaped isodensity contours or negative values.
These problems become more acute as the ellipticity increases, and the approach is generally unsuited for axis ratios of $q \approx 0.5$ and below.

More realistic lens models might therefore be created from elliptical surface mass densities.
The properties of such lenses were described early by \citet*{1973ApJ...185..747B} and \citet*{1975ApJ...195...13B}, who introduced a complex formalism of gravitational lensing to simplify the necessary calculations.
Later \citet{1990A&A...231...19S} found the equivalent deflection angle in terms of two-dimensional real coordinates.
However, due to the loss of symmetry, it is often no longer possible to find the lens properties of elliptical mass distributions analytically.
Notable exceptions are the singular and nonsingular isothermal ellipsoids described by \citet*{1993ApJ...417..450K} and \citet*{1994A&A...284..285K} using the complex formalism, which feature beautifully simple closed-form solutions while still being widely applicable.
Despite the success of this model, real observations often require more flexibility in the assumed lens profile, and no equally well-established elliptical lens with a simple mathematical form and numerical implementation is available today.

In the present article, a general class of lenses following an elliptical power law profile is investigated.
Previously, a number of numerical recipes for the nonsingular variant of this model have been presented \citep{1991A&A...247..269S,1998ApJ...502..531B}.
In section~\ref{sec:powerlaw}, the elliptical power law profile lens is reexamined using the complex formalism of lensing, with the goal of finding compact expressions for the deflection, potential, shear, and magnification.
In section~\ref{sec:special}, it is shown that the new expressions reduce to known results for a number of special cases of the power law profile.
Finally, in section~\ref{sec:numerics} it is demonstrated how the results naturally lead to a simple and efficient scheme for numerical computation.

\section{The elliptical power law profile lens}
\label{sec:powerlaw}

The dimensionless surface mass density $\kappa$ of a lens following a power law profile with circular symmetry is (e.g. \citealp{1998MNRAS.296..800E}, \citealp{2006glsw.conf.....M} and references therein)%
\begin{equation}\label{eq:kappa-r}
	\kappa(r) = \frac{2 - t}{2} \left(\frac{b}{r}\right)^{t} \;,
\end{equation}
where $0 < t < 2$ is the slope of the profile, $b > 0$ is the scale length and $r > 0$ is the distance from the centre of the mass distribution.
Such a profile arises from a spherically symmetric three-dimensional mass distribution $\rho(r) \propto r^{-t-1}$.
The power law profile lens is a versatile model; it contains as
special cases the singular isothermal sphere for $t = 1$, the point
mass for $t = 2$, and approximates the inner slope of the \citet*{1996ApJ...462..563N} and \citet{1999MNRAS.310.1147M} profiles for $t = 0$ and $t = 1/2$, respectively.

In order to turn the profile~\eqref{eq:kappa} into an elliptical surface mass distribution, the $x$-axis is now stretched by a factor of $q^{-1}$, where $0 < q \leq 1$ is a constant parameter of the new profile.
It is clear that after this operation, the formerly circular isodensity contours $r = \mathrm{const}$ have indeed become ellipses with semi-major axis $r/q$, semi-minor axis $r$, and axis ratio $q$.
The resulting elliptical power law profile can be written as
\begin{equation}\label{eq:kappa}
	\kappa(R) = \frac{2 - t}{2} \left(\frac{b}{R}\right)^{t} \;,
\end{equation}
where $R$ is the \emph{elliptical radius} defined by
\begin{equation}\label{eq:R}
	R = \sqrt{q^2 x^2 + y^2} \;,
\end{equation}
i.e. the semi-minor axis of the ellipse passing through $x,y$.
One also defines\footnote{%
This definition uses the two-argument inverse tangent $\arctan(x,y)$, which respects the quadrant of $x,y$ to return the correct angle.
}
the corresponding \emph{elliptical angle}
\begin{equation}\label{eq:varphi}
	\varphi = \arctan(q x, y)
\end{equation}
as the polar angle of $x,y$ when transforming back to circular symmetry.
The inverse coordinate transformation is
\begin{equation}\label{eq:tfm}
\begin{split}
	x &= R/q \, \cos(\varphi) \;, \\
	y &= R \, \sin(\varphi) \;,
\end{split}
\end{equation}
where $R$ and $\varphi$ are limited to $R > 0$ and $\varphi \in [0,2\pi]$, respectively.

For the following calculation, the complex formulation of lensing for spheroidal mass distributions is used, as laid out by \citet{1973ApJ...185..747B} and \citet{1975ApJ...195...13B}, with contributions from \citet{1984MNRAS.208..511B}, and \citet{1990A&A...231...19S}, who also investigated the potential of such lenses.
These results were later used to great effect by \citet{1993ApJ...417..450K} and even more so \citet{1994A&A...284..285K} in the analysis of the singular and nonsingular isothermal ellipsoid.
The derivations of this section closely follow their outlined path.

In complex notation, the gravitational lens equation is
\begin{equation}
	z' = z - \alpha(z) \;,
\end{equation}
where the complex coordinate $z = x + \I y$ and complex deflection angle $\alpha = \alpha_x + \I \alpha_y$ replace the corresponding two-dimensional vector quantities.
Following \citet{1990A&A...231...19S}, one can further introduce the complex deflection potential $\psi(z)$, the real part of which is the usual deflection potential of the real formulation.
It is related to the complex deflection angle $\alpha(z)$ by the Wirtinger derivative (see e.g. \citealp{1994A&A...284..285K} for a definition)
\begin{equation}\label{eq:psi-alpha}
	\co\alpha(z) = 2 \, \dfrac{\partial\psi}{\partial z} \;,
\end{equation}
where the asterisk $\co\alpha$ denotes complex conjugation.
Similarly, such relations can also be found for the second derivatives of the potential, namely the complex shear $\gamma = \gamma_1 + \I \gamma_2$, given by
\begin{equation}\label{eq:gamma-alpha}
	\co\gamma(z) = \dfrac{\partial \co\alpha}{\partial z} \;,
\end{equation}
and the convergence
\begin{equation}\label{eq:kappa-alpha}
	\kappa(z) = \dfrac{\partial \co\alpha}{\partial \co{z}} \;.
\end{equation}
Since the latter expression must recover the surface mass density, it presents a direct check for the correctness of the calculations.

The complex deflection angle for an elliptical, homoeoidal mass distribution was given by \cite{1975ApJ...195...13B}.
Using the elliptical radius $R$ defined in~\eqref{eq:R}, the deflection angle of an elliptical surface mass density $\kappa(R)$ is
\begin{equation}\label{eq:deflection}
	\co\alpha(z) = 2 \, \frac{\sqrt{\,z^2\,}}{z} \int_0^{R(z)} \! \D r \, \frac{\kappa(r) \, r}{\sqrt{q^2 \, z^2 - (1-q^2) \, r^2}} \;,
\end{equation}
where $R(z)$ is the semi-minor axis of the ellipse passing through point~$z$.
The factor in front of the integral is due to \citet{1984MNRAS.208..511B}; it ensures the correct sign of the deflection in all quadrants of the complex plane.
Inserting the elliptical power law profile~\eqref{eq:kappa} into integral~\eqref{eq:deflection}, the complex deflection angle can be written as
\begin{equation}
\begin{split}
	\co\alpha(z)
	&= \frac{2-t}{q} \, \frac{b^t}{z} \, \int_0^{R(z)} \! \D r \, r^{1-t} \, \left(1 - \frac{1-q^2}{q^2} \, \frac{r^2}{z^2}\right)^{-1/2} \\
	&= \frac{2-t}{2 q} \, \frac{b^2}{z} \, \left(\frac{b}{R(z)}\right)^{t-2} \int_0^1 \! \D\xi \, \xi^{-t/2} \, \left(1 - \frac{1-q^2}{q^2} \, \frac{R(z)^2}{z^2} \, \xi\right)^{-1/2} \\
	&= \frac{1}{q} \, \frac{b^2}{z} \, \left(\frac{b}{R(z)}\right)^{t-2} {}_2F_1\left(\tfrac{1}{2}, 1 - \tfrac{t}{2}; 2 - \tfrac{t}{2}; \tfrac{1 - q^2}{q^2} \, \tfrac{R(z)^2}{z^2}\right) \;,
\end{split}
\end{equation}
where the change of variable $r \to \xi = r^2/R(z)^2$ was used in the first step, and the integral representation of the Gaussian hypergeometric function ${}_2F_1(a, b; c; z)$ was used in the second step.
Because its parameters are related as $c = a + b + 1/2$, a quadratic transformation of the hypergeometric function exists, and the complex deflection angle can further be simplified to
\begin{equation}\label{eq:coalpha}
	\co\alpha(R,\varphi)
	= \frac{2\, b}{1 + q} \left(\frac{b}{R}\right)^{t-1} \E^{-\I \varphi} \, {}_2F_1\left(1, \tfrac{t}{2}; 2 - \tfrac{t}{2}; -\tfrac{1-q}{1+q} \, \E^{-\I 2\varphi}\right) \;,
\end{equation}
where coordinate transformation~\eqref{eq:tfm} was applied to the complex variable $z = x + \I y$.
For reference, the complex deflection angle before complex conjugation is
\begin{equation}\label{eq:alpha}
	\alpha(R,\varphi)
	= \frac{2\, b}{1 + q} \left(\frac{b}{R}\right)^{t-1} \E^{\I \varphi} \, {}_2F_1\left(1, \tfrac{t}{2}; 2 - \tfrac{t}{2}; -\tfrac{1-q}{1+q} \, \E^{\I 2\varphi}\right) \;.
\end{equation}
This result is a beautiful factorisation of $\alpha$ into its (elliptical) radial and angular parts; a fact that will be exploited in section~\ref{sec:numerics} to calculate the deflection quickly, in spite of the hypergeometric function it contains.

Taking inspiration from the singular isothermal ellipsoid, one finds that a potential which solves equation~\eqref{eq:psi-alpha} for complex deflection~\eqref{eq:coalpha} is given by
\begin{equation}\label{eq:psi}
	\psi(z) = \frac{1}{2-t} \frac{z \, \co\alpha(z) + \co{z} \, \alpha(z)}{2} \;.
\end{equation}
In real coordinates, this expression reduces to the more familiar
\begin{equation}\label{eq:psi-xy}
	\psi(x,y) = \frac{x \, \alpha_x + y \, \alpha_y}{2-t} \;.
\end{equation}
It is clear that this potential can be calculated at very low cost if the deflection angle~$\alpha$ has already been found.
Checking the result is tedious; one possibility is to substitute $\E^{-\I \varphi} = (q x - \I y)/R$ and $\E^{-\I 2\varphi} = (q x - \I y)/(q x + \I y)$ in deflection~\eqref{eq:coalpha} and express the Wirtinger derivative in \eqref{eq:psi-alpha} in terms of real coordinates as $\partial/\partial z = (\partial/\partial x - \I \partial/\partial y)/2$.

\begin{figure*}
\centering%
\includegraphics[width=\textwidth]{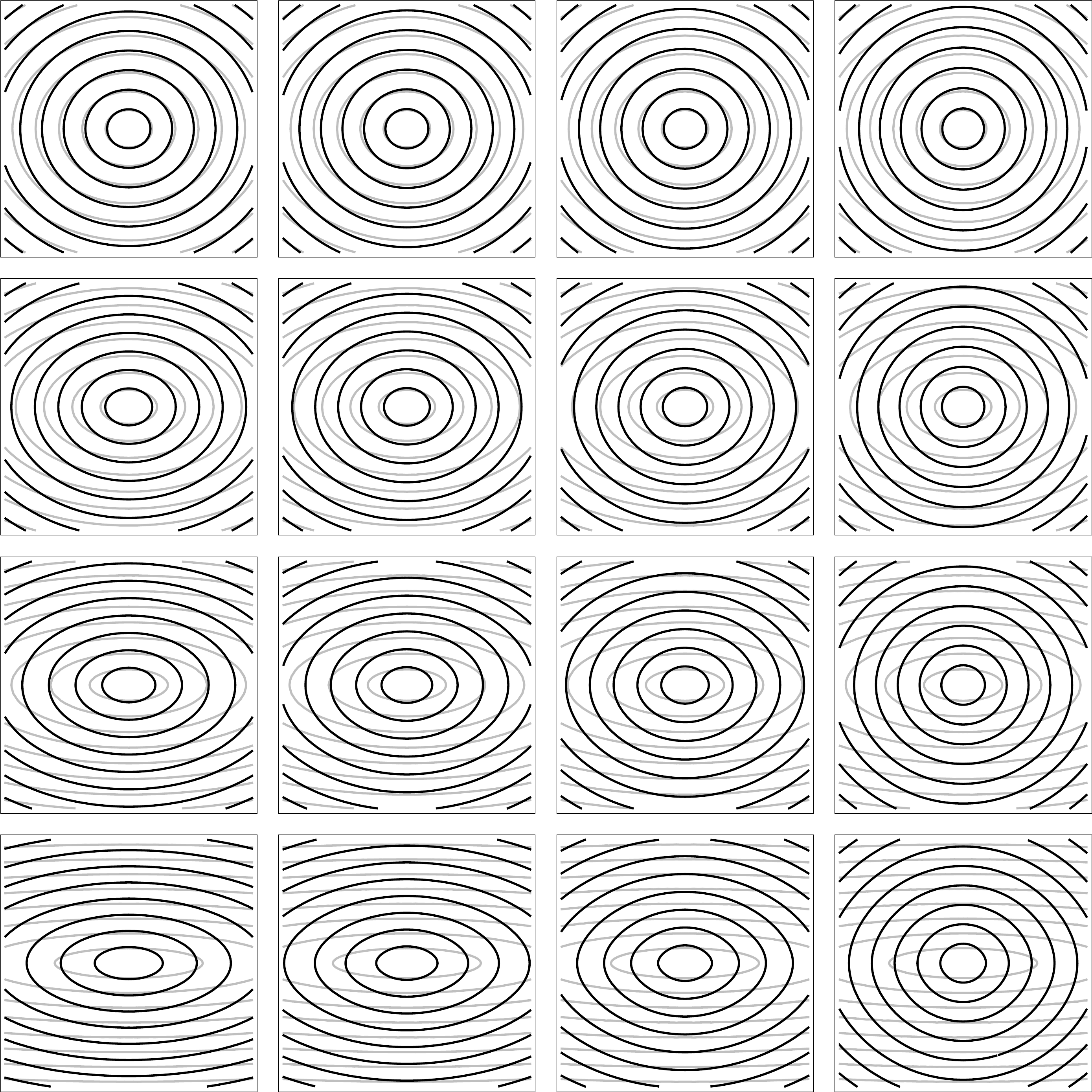}%
\caption{%
Isocontours of the deflection potential $\psi$ (black) and surface mass density $\kappa$ (grey) for an elliptical power law profile lens in physical coordinates $x,y$.
For the purpose of illustration, the contour levels are equally spaced along the diagonals.
The slope of the power law profile varies from $t = 0.25$ (left) to $t = 1.75$ (right) in steps of $0.5$.
The axis ratio varies from $q = 0.8$ (top) to $q = 0.2$ (bottom) in steps of $0.2$.
}%
\label{fig:psi}%
\end{figure*}

The complex potential~\eqref{eq:psi} evidently has no imaginary component.
In its form~\eqref{eq:psi-xy}, it can therefore be taken to be the usual real deflection potential satisfying $\vec\alpha = \nabla\psi$ \citep{1990A&A...231...19S}.
Fig.~\ref{fig:psi} shows the potential for various settings of the power law slope $t$ and axis ratio $q$.
It has been noted that the potential is always "rounder", i.e. less eccentric, than the surface mass density, and this effect is clearly visible.

Calculating the shear from equation~\eqref{eq:gamma-alpha} is straightforward, and results in
\begin{equation}\label{eq:gamma}
	\co\gamma(z) = - \kappa(z) \, \frac{\co{z}}{z} + (1 - t) \, \frac{\co\alpha(z)}{z} \;.
\end{equation}
For computations, the shear is more clearly expressed in terms of the physical polar coordinates $r$ and $\theta$ as
\begin{equation}
	\gamma = - \E^{\I 2\theta} \, \kappa + \left(1 - t\right) \E^{\I \theta} \, \frac{\alpha}{r} \;,
\end{equation}
once more recalling the singular isothermal ellipsoid ($t = 1$).
As in the case of the potential, the shear is readily calculated once the surface mass density $\kappa$ and complex deflection angle $\alpha$ are known.

\begin{figure*}
\centering%
\includegraphics[width=\textwidth]{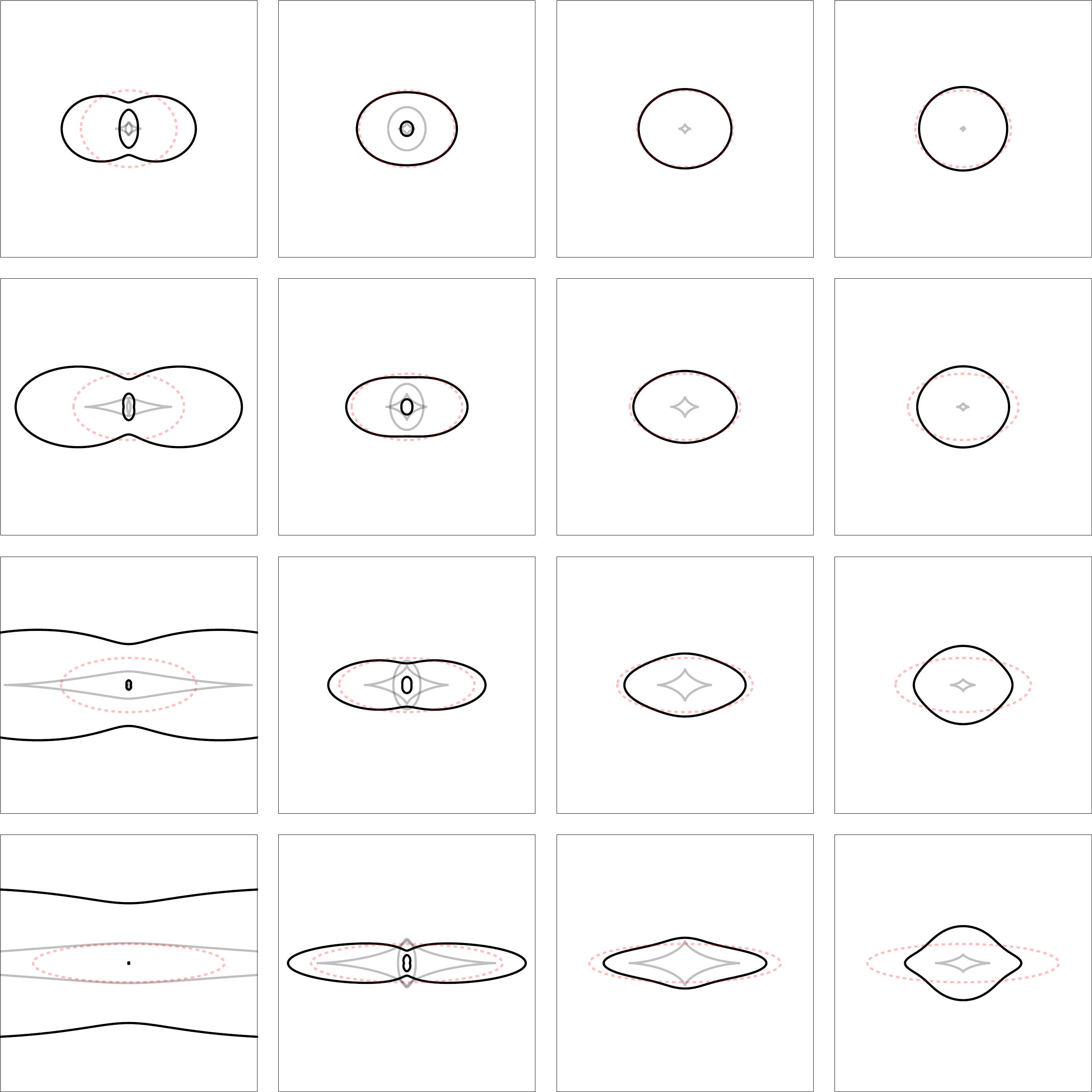}%
\caption{%
Critical lines (black) and caustics (grey) of the elliptical power law profile lens.
Also shown is an ellipse with semi-minor axis equal to the scale length $b$ and axis ratio $q$ (red, dotted).
For the purpose of illustration, the scale length $b$ decreases as $q^{1/2}$.
The slope of the power law profile varies from $t = 0.25$ (left) to $t = 1.75$ (right) in steps of $0.5$.
The axis ratio varies from $q = 0.8$ (top) to $q = 0.2$ (bottom) in steps of $0.2$.
}%
\label{fig:crit}%
\end{figure*}

Given shear $\gamma$ and convergence $\kappa$, it is now possible to find the (inverse) magnification $\mu^{-1} = (1 - \kappa)^2 - |\gamma|^2$ of the elliptical power law profile lens, which is
\begin{equation}\label{eq:mu}
	\mu^{-1} = 1 - 2\kappa \left(1 - (1 - t) \, \frac{x \, \alpha_x + y \, \alpha_y}{r^2}\right) - (1 - t)^2 \, \frac{|\alpha|^{2}}{r^2} \;.
\end{equation}
Here $r$ denotes the radius $r^2 = x^2 + y^2$ in physical coordinates.
Using expression~\eqref{eq:mu}, it is possible to determine the critical lines $\mu^{-1} = 0$ and corresponding caustics of the lens.
Because the resulting equations contain the hypergeometric function, their solutions are found numerically.
The critical lines and caustics for a number of settings of the power law slope $t$ and axis ratio $q$ are shown in Fig.~\ref{fig:crit}.

\section{Special cases}
\label{sec:special}

In order to check the presented solutions for the deflection~\eqref{eq:alpha}, potential~\eqref{eq:psi}, shear~\eqref{eq:gamma}, and magnification~\eqref{eq:mu}, it is useful to compare them to known special cases of the power law slope $t$.

The singular isothermal ellipsoid is a power law profile lens with slope $t = 1$.
The hypergeometric function can be simplified in this case, yielding the complex deflection angle in physical coordinates
\begin{equation}
	\alpha(x,y)
	= \frac{2 b}{\sqrt{1 - q^2}} \arctan\left(\frac{\sqrt{1 - q}}{\sqrt{1 + q}} \frac{\sqrt{q x + \I y}}{\sqrt{q x - \I y}}\right) \;.
\end{equation}
Taking the real and imaginary part and using the sum formula for the inverse tangent reduces the expression to the commonly used form as reported e.g. by \cite{2006glsw.conf.....M}.
It is trivial to check the potential $\psi = x \, \alpha_x + y \, \alpha_y$, shear $\gamma = -\kappa \, \co{z}/z$, and magnification $\mu^{-1} = 1 - 2\kappa$.

The limit $t \to 2$ takes the power law profile to a point mass.
The hypergeometric function exists in the limit, resulting in the complex deflection angle
\begin{equation}
	\alpha(x,y) = \frac{b^2 \, (x + \I y)}{q \, r^2} \;.
\end{equation}
As expected for a point mass, the deflection is always circularly symmetric (note $r$ instead of $R$), and the axis ratio $q$ appears only in the form of a constant of normalisation, making the Einstein radius of the lens $r_E = b \, q^{-1/2}$.
While the potential~\eqref{eq:psi} cannot be evaluated in the limit $t \to 2$, both the shear $\gamma = - b^2/(q r^2) \, \E^{\I 2\theta}$ and the magnification $\mu^{-1} = 1 - b^4/(q^2 r^4)$ are easily checked and agree with the known results for the corrected Einstein radius. 

The final case of interest is the uniform critical mass sheet in the limit $t \to 0$.
In this case, the deflection becomes
\begin{equation}\label{eq:t0}
	\alpha(R, \varphi) = \frac{2}{1 + q} \, R \, \E^{\I \varphi} \;,
\end{equation}
which has a clearly unphysical dependency on $\varphi$, as the surface mass density $\kappa \equiv 1$ has circular symmetry.
However, the same result~\eqref{eq:t0} is obtained if $\kappa \equiv 1$ is inserted into deflection~\eqref{eq:deflection} from \citet{1975ApJ...195...13B} directly.
The problematic result is therefore a limitation of the formalism for elliptical mass distributions, and not the elliptical power law profile lens.

\section{Numerical evaluation}
\label{sec:numerics}

\begin{figure}[t]
\centering%
\includegraphics[width=\columnwidth]{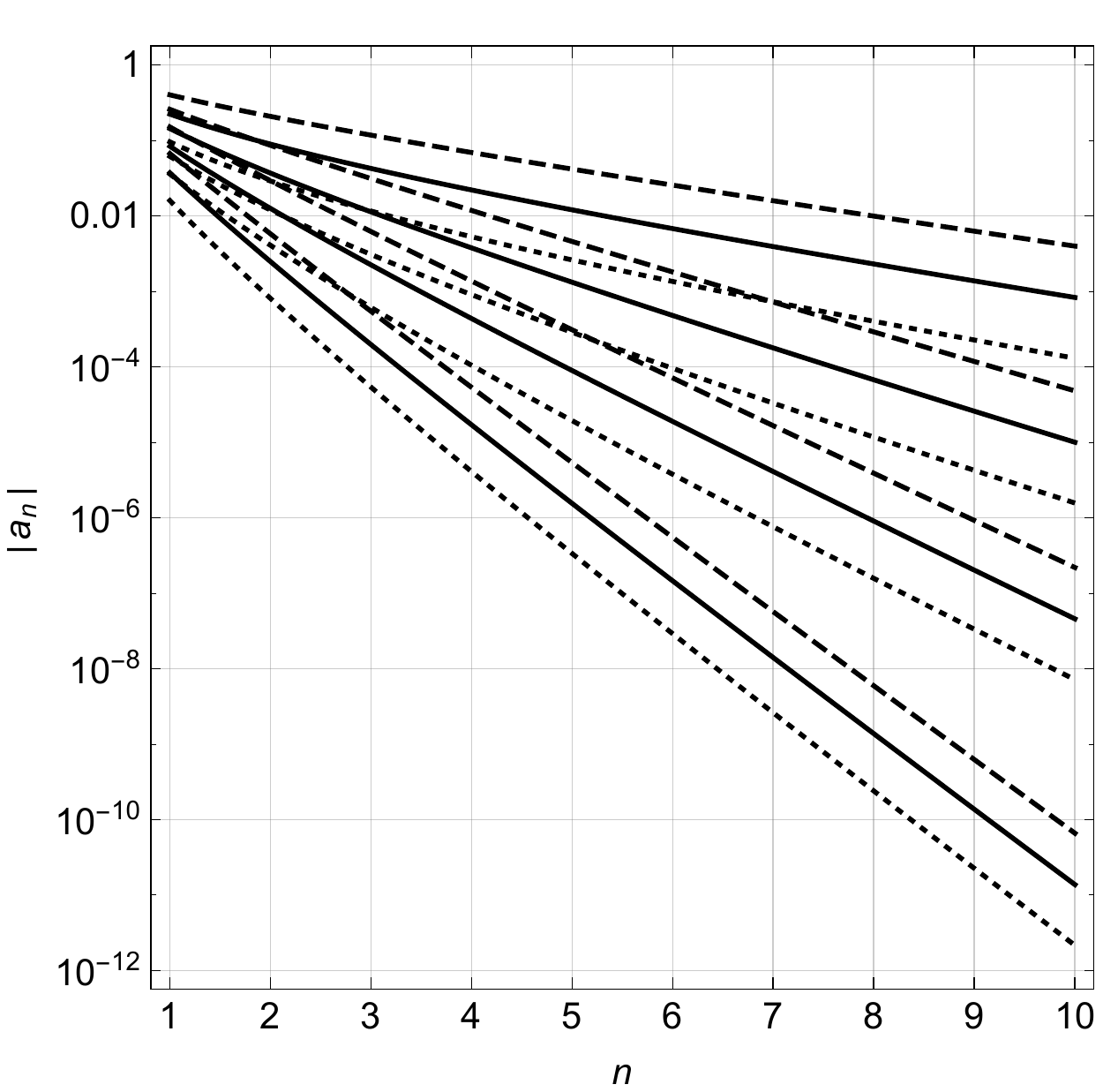}%
\caption{%
Absolute value of the series coefficients $a_n$ for the elliptical power law profile lens.
Shown are different values $t = 0.5$ (dotted), $t = 1.0$ (solid), and $t = 1.5$ (dashed) for the slope of the power law and axis ratios from $q = 0.8$ (bottom) to $q = 0.2$ (top) in steps of $0.2$.
}%
\label{fig:a_n}%
\end{figure}

The deflection angle~\eqref{eq:alpha} is at the heart of the elliptical power law profile lens, as the potential~\eqref{eq:psi} and shear~\eqref{eq:gamma} can all be expressed in terms of it.
It is therefore necessary to find a fast method for evaluating the
contained hypergeometric function to make this a useful model for
numerical simulation and modeling purposes.
For this, it is useful to separate the radial and angular parts of the complex deflection angle~\eqref{eq:alpha}, which becomes
\begin{equation}
	\alpha(R,\varphi)
	= \frac{2\, b}{1 + q} \left(\frac{b}{R}\right)^{t-1} \omega(\varphi) \;.
\end{equation}
The angular dependency of $\alpha$ is contained in the function
\begin{equation}\label{eq:omega}
	\omega(\varphi)
	= \E^{\I \varphi} \, {}_2F_1\left(1, \tfrac{t}{2}; 2 - \tfrac{t}{2}; -f \, \E^{\I 2\varphi}\right) \;,
\end{equation}
where $f = \frac{1-q}{1+q}$ is the second flattening of an ellipse with axis ratio $q$.
Because $0 < q \leq 1$, the range of $f$ is $0 \leq f < 1$, and it follows that the hypergeometric function in expression~\eqref{eq:omega} has a convergent series representation
\begin{equation}
	\omega(\varphi)
	= \sum_{n=0}^{\infty} \frac{\Gamma\big(2-\frac{t}{2}\big) \, \Gamma\big(n+\frac{t}{2}\big)}{\Gamma\big(\frac{t}{2}\big) \, \Gamma\big(n+2-\frac{t}{2}\big)} \, (-f)^n \, \E^{\I (2n + 1) \, \varphi} \;.
\end{equation}
This, on the other hand, is nothing but a Fourier-type series
\begin{equation}\label{eq:omega-fourier}
	\omega(\varphi)
	= \sum_{n=0}^{\infty} a_n \, \E^{\I (2n + 1) \, \varphi}
\end{equation}
containing only positive and odd terms $2n+1$, the coefficients of which are given by
\begin{equation}
	a_n
	= \frac{\Gamma\big(2-\frac{t}{2}\big) \, \Gamma\big(n+\frac{t}{2}\big)}{\Gamma\big(\frac{t}{2}\big) \, \Gamma\big(n+2-\frac{t}{2}\big)} \, (-f)^n \;.
\end{equation}
Since this is a hypergeometric series, the ratio of two subsequent series coefficients is simple,
\begin{equation}\label{eq:coeff-ratio}
	\frac{a_n}{a_{n-1}} = -f \, \frac{2n - (2 - t)}{2n + (2 - t)} \;.
\end{equation}
It is clear that the magnitude of the series terms drops off almost geometrically, with an asymptotic rate of $f$.
This behaviour is shown in Fig.~\ref{fig:a_n} for various settings of the power law slope $t$ and axis ratio $q$.

The ratio \eqref{eq:coeff-ratio} can be used to iteratively calculate the terms of series~\eqref{eq:omega-fourier}.
Introducing symbols for the summands as
\begin{equation}\label{eq:omega-series}
	\omega = \sum_{n=0}^{\infty} \Omega_n \;,
\end{equation}
the $n$-th term $\Omega_n$ is related to the previous one as
\begin{equation}\label{eq:Omega}
	\Omega_n = - f \, \frac{2n - (2 - t)}{2n + (2 - t)} \,\E^{\I 2\varphi} \, \Omega_{n-1}\;.
\end{equation}
This reduces the calculation of the deflection to summation and (complex) multiplication.
Furthermore, since the computation is iterative, it can be continued easily until the desired precision or accuracy in the deflection is reached.

Instead of the numerical scheme given by equations~\eqref{eq:omega-series} and~\eqref{eq:Omega}, it can be advantageous to forego the use of complex numbers, particularly in computer implementations.
Just as the two-dimensional deflection angle~$\vec\alpha$ is given by the real and imaginary components of the complex deflection angle~$\alpha$, the complex function $\omega$ for the angular dependency corresponds to a two-dimensional vector $\vec\omega$ in the real formulation.
Expanding the exponential in Fourier series~\eqref{eq:omega-fourier} into its real and imaginary parts, the components of $\vec\omega$ can be written as the individual series
\begin{equation}
\begin{split}\label{eq:omega-xy}
	\omega_x(\varphi)
	&= \sum_{n=0}^{\infty} a_n \, \cos\left((2n + 1) \, \varphi\right) \;, \\
	\omega_y(\varphi)
	&= \sum_{n=0}^{\infty} a_n \, \sin\left((2n + 1) \, \varphi\right) \;.
\end{split}
\end{equation}
Just as in the complex case, the real components of the deflection angle can be calculated iteratively.
Writing
\begin{equation}
\begin{split}	
	\omega_x &= \sum_{n=0}^{\infty} \Omega_{x,n} \;, \\
	\omega_y &= \sum_{n=0}^{\infty} \Omega_{y,n} \;,
\end{split}
\end{equation}
the relations for the $n$-th terms can be found by understanding the complex product $\E^{\I 2\varphi} \, \Omega_{n-1}$ in expression~\eqref{eq:Omega} as the two-dimensional matrix multiplication
\begin{equation}\label{eq:Omega2d}
	\vec\Omega_n = - f \, \frac{2n - (2 - t)}{2n + (2 - t)} \, \mat{R}_{2\varphi} \, \vec\Omega_{n-1} \;,
\end{equation}
where $\mat{R}_{2\varphi}$ is the rotation matrix.
In components, this is
\begin{equation}
\begin{split}
	\Omega_{x,n} &= - f \, \frac{2n - (2 - t)}{2n + (2 - t)} \left( \cos(2\varphi) \, \Omega_{x,n-1} - \sin(2\varphi) \, \Omega_{y,n-1} \right) \;, \\
	\Omega_{y,n} &= - f \, \frac{2n - (2 - t)}{2n + (2 - t)} \left( \sin(2\varphi) \, \Omega_{x,n-1} + \cos(2\varphi) \, \Omega_{y,n-1} \right) \;.
\end{split}
\end{equation}
This prescription can make evaluation of the deflection suitably fast to use the elliptical power law profile lens for ray-tracing in model fitting and Monte Carlo inference, where a large number of calculations are necessary.

\section{Conclusion}

It was shown how the power law profile~\eqref{eq:kappa-r} for lensing can be turned into a surface mass density~\eqref{eq:kappa} with elliptical isodensity contours.
Following the groundbreaking work by \citet{1973ApJ...185..747B} and \citet{1975ApJ...195...13B} on ellipsoidal lenses, and \citet{1994A&A...284..285K} on the singular isothermal ellipsoid, the properties of the elliptical power law profile lens have been derived using the complex formalism of gravitational lensing.
It was found that the deflection angle~\eqref{eq:alpha} can only be expressed in terms of the Gaussian hypergeometric function ${}_2F_1(a,b;c;z)$, thereby severely limiting the analytical tractability of the lens equation.
Further results for the potential~\eqref{eq:psi}, the shear~\eqref{eq:gamma}, and the magnification~\eqref{eq:mu} could all be given in terms of the complex deflection angle, which therefore becomes the central quantity for computations.
These results have been checked in section~\ref{sec:special} using a number of known special and limiting cases of the power law profile.
Finally, a recipe was given in section~\ref{sec:numerics} for the quick numerical evaluation of the deflection angle, and consequently other lensing quantities.

\begin{acknowledgements}
This research is part of project GLENCO, funded under the European Seventh Framework Programme, Ideas, Grant Agreement n. 259349.
\end{acknowledgements}

\bibliographystyle{aa}
\bibliography{powerlaw}

\end{document}